\documentclass[final]{aipproc}
\usepackage{graphics,graphicx,amsmath,array,hhline,fancyhdr,mathptm}
\usepackage[english]{babel}

\layoutstyle{6x9}

\begin{document}

\title{Understanding the rapidity dependence of the elliptic flow and the HBT radii
      at RHIC}


\author{M. Csan\'ad}
{address={Department of Atomic Physics, ELTE, Budapest, P\'azm\'any P. 1/A,
H-1117}}

\author{T. Cs\"org\H{o}}
{address={MTA KFKI RMKI, H - 1525 Budapest 114, P.O.Box 49,
Hungary}}

\author{B. L\"orstad}
{address={Department of Physics, University of Lund, S-22362 Lund,
Sweden} }

\author{A. Ster}
{address={MTA KFKI RMKI, H - 1525 Budapest 114, P.O.Box 49,
Hungary}}

\begin{abstract}
The pseudo-rapidity dependence of the elliptic flow at various
excitation energies measured by the PHOBOS Collaboration in Au+Au
collisions at RHIC is one of the surprising results that has not
been explained before in terms of hydrodynamical models. Here we
show that these data are in agreement with theoretical predictions and
satisfy the universal scaling relation predicted by
the Buda-Lund hydrodynamical model, based on exact solutions of
perfect fluid hydrodynamics.
We also show a theoretical prediction on the rapidity and transverse momentum
scaling of the HBT radii measured in heavy ion collisions, based on the Buda-Lund model.
\end{abstract}

\maketitle
\section{Introduction}

One of the unexpected results from experiments at the Relativistic
Heavy Ion Collider (RHIC) is the relatively strong second harmonic
moment of the transverse momentum distribution, referred to as the
elliptic flow. Measurements of the elliptic flow by the PHENIX,
PHOBOS and STAR collaborations (see
refs.~\cite{Back:2004zg,Adler:2003kt,Adler:2001nb,Sorensen:2003wi})
reveal rich details in terms of its dependence on particle type,
transverse and longitudinal momentum variables, on the centrality
and the bombarding energy of the collision. In the soft transverse
momentum region, these measurements at mid-rapidity are reasonably
well described by hydrodynamical
models~\cite{Adcox:2004mh,Adams:2005dq}.
However, the dependence
of the elliptic flow on the longitudinal momentum variable
pseudo-rapidity and its excitation function has resisted
descriptions in terms of hydrodynamical models (but see their new description
by the SPHERIO model~\cite{Grassi:2005pm}).

Here we show that these data are consistent with the theoretical
and analytic predictions that are based on eqs. (1-6) of ref.~\cite{hidro3},
that is, on perfect fluid hydrodynamics.

We furthermore calculate rapidity dependent HBT (Bose-Einstein) radii in the framework of the model
and make prediction on the universal scaling of these observables.

Our tool in describing the pseudorapidity-dependent elliptic flow and HBT radii
is the Buda-Lund hydrodynamical model. The Buda-Lund hydro model~\cite{Csorgo:1995bi} is successful
in describing the BRAHMS, PHENIX, PHOBOS
and STAR data on identified single particle spectra and the
transverse mass dependent Bose-Einstein or HBT radii as well as
the pseudorapidity distribution of charged particles in Au + Au
collisions both at $\sqrt{s_{\rm{NN}}} = 130 $
GeV~\cite{ster-ismd03} and at $\sqrt{s_{\rm{NN}}} = 200 $
GeV~\cite{mate-warsaw03}. However the elliptic flow would be zero
in an axially symmetric case, so we developed the ellipsoidal
generalization of the model that describes an expanding ellipsoid
with principal axes $X$, $Y$ and $Z$. Their derivatives with
respect to proper-time (expansion rates) are denoted by $\dot X$,
$\dot Y$ and  $\dot Z$.

The generalization goes back to the original one, if the
transverse directed principal axes of the ellipsoid are equal, ie
$X=Y$ (and also $\dot X=\dot Y$).

The deviation from axial symmetry can be measured by the momentum-space eccentricity,
\begin{equation}
\epsilon_p = \frac{\dot X^2 - \dot Y^2}{\dot X^2 + \dot Y^2}.
\end{equation}

The exact analytic solutions of hydrodynamics (see
ref.~\cite{hidro3,hidro1,hidro2}), which form the basis of the
Buda-Lund hydro model, develop Hubble-flow for late times, ie $X
\rightarrow_{\tau \rightarrow \infty} \dot X \tau$, so the
momentum-space eccentricity $\epsilon_p$ nearly equals space-time
eccentricity $\epsilon$.

Let us introduce $\Delta\,\eta$ additionally. It represents the
elongation of the source expressed in units of space-time rapidity.
Let us consider furthermore that at the freeze-out $\tau\,\Delta \eta = Z$ and
$Z \approx \dot Z\,\tau$, and so $\Delta \eta \approx \dot Z$

Hence, in this paper we extract
space-time eccentricity ($\epsilon$), average transverse flow
($u_t$) and longitudinal elongation ($\Delta \eta$) from the data,
instead of $\dot X$, $\dot Y$ and $\dot Z$.

In the time dependent hydrodynamical solutions, these values
evolve in time, however, it was show in ref.~\cite{hidro4} that
$\dot X$, $\dot Y$ and $\dot Z$, and so $\epsilon$, $u_t$ and $\Delta \eta$
become constants of the motion in the late stages of the expansion.

\section{Rapidity dependent elliptic flow}

The result for the elliptic flow (under certain conditions
detailed in ref~\cite{bl-ell}) is the following simple universal
scaling law:
\begin{equation}
v_2=\frac{I_1(w)}{I_0(w)}.\label{e:v2w}
\end{equation}

The model predicts an \emph{universal scaling:} every $v_2$
measurement is predicted to fall on the same \emph{universal}
scaling curve $I_1/I_0$ when plotted against $w$.

This means, that $v_2$ depends on any physical parameter
(transverse or longitudinal momentum, center of mass energy,
centrality, type of the colliding nucleus etc.) only through the (universal)
scaling paremeter $w$.

Here $w$ is the scaling variable, defined by
\begin{equation}w=\frac{p_t^2}{4
   \overline{m}_t} \left(\frac{1}{T_{*,y}}
   -\frac{1}{T_{*,x}}\right),
\end{equation}
and
\begin{eqnarray}
   T_{*,x}&=&T_0+\overline{m}_t \, \dot X^2
       \frac{T_0}{T_0 +\overline{m}_t a^2},\\
   T_{*,y}&=&T_0+\overline{m}_t \, \dot Y^2
     \frac{T_0}{T_0 +\overline{m}_t a^2},
\end{eqnarray}
  and
\begin{equation}
    \overline{m}_t = m_t \cosh(\eta_{s}-y).
\end{equation}
Here $a=\langle \Delta T/T \rangle_t$ measures the temperature
gradient in the transverse direction, at the freeze-out, $m_t$ is
the transverse mass, $T_0$ the central temperature at the
freeze-out, while $\eta_{s}$ is the space-time rapidity of the
saddle-point (point of maximal emittivity). This saddlepoint
depends on the rapidity, the longitudinal expansion, the
transverse mass and on the central freeze-out temperature:
\begin{equation}
\eta_{s}-y = \frac{y}{1+\Delta \eta \frac{m_t}{T_0}},
\end{equation}
where $y = 0.5 \log(\frac{E + p_z}{E - p_z})$ is the rapidity.

More details about the ellipsoidally symmetric model and its result on
$v_2(\eta)$ can be found in ref.~\cite{bl-ell}.

Eq.~\ref{e:v2w} depends, for a given centrality class, on rapidity
$y$ and transverse mass $m_t$. Before comparing our result to the
$v_2(\eta)$ data of PHOBOS, we thus performed a saddle point
integration in the transverse momentum variable
and performed a change of variables to the pseudo-rapidity
$\eta=0.5 \log(\frac{|p| + p_z}{|p| - p_z})$,
similarly to ref.~\cite{Kharzeev:2001gp}. This way, we have evaluated the
single-particle invariant spectra in terms of the
variables $\eta$ and $\phi$, and calculated $v_2(\eta)$
from this distribution, a procedure corresponding to
the PHOBOS measurement decribed in ref.~\cite{Back:2004zg}.

We have found that the essential fit parameters are $\epsilon$ and
$\Delta \eta$, and the quality of the fit is insensitive to the
precise value of $T_0$, $a$ and $u_t$. These parameters
dominate the azimuthal-averaged single particle spectra as well as
the HBT (Bose-Einstein) radii, however they only marginally
influence $v_2$. Their precise value is
irrelevant in a broad region of values and does not influence the confidence level of the
$v_2(\eta)$ fits. Hence we have fixed their values as given in the
caption of table~\ref{t:fit}. We also excluded points with large
rapidity from the fits in case of lower center of mass energies.

\begin{figure}
  \includegraphics[width=250pt]{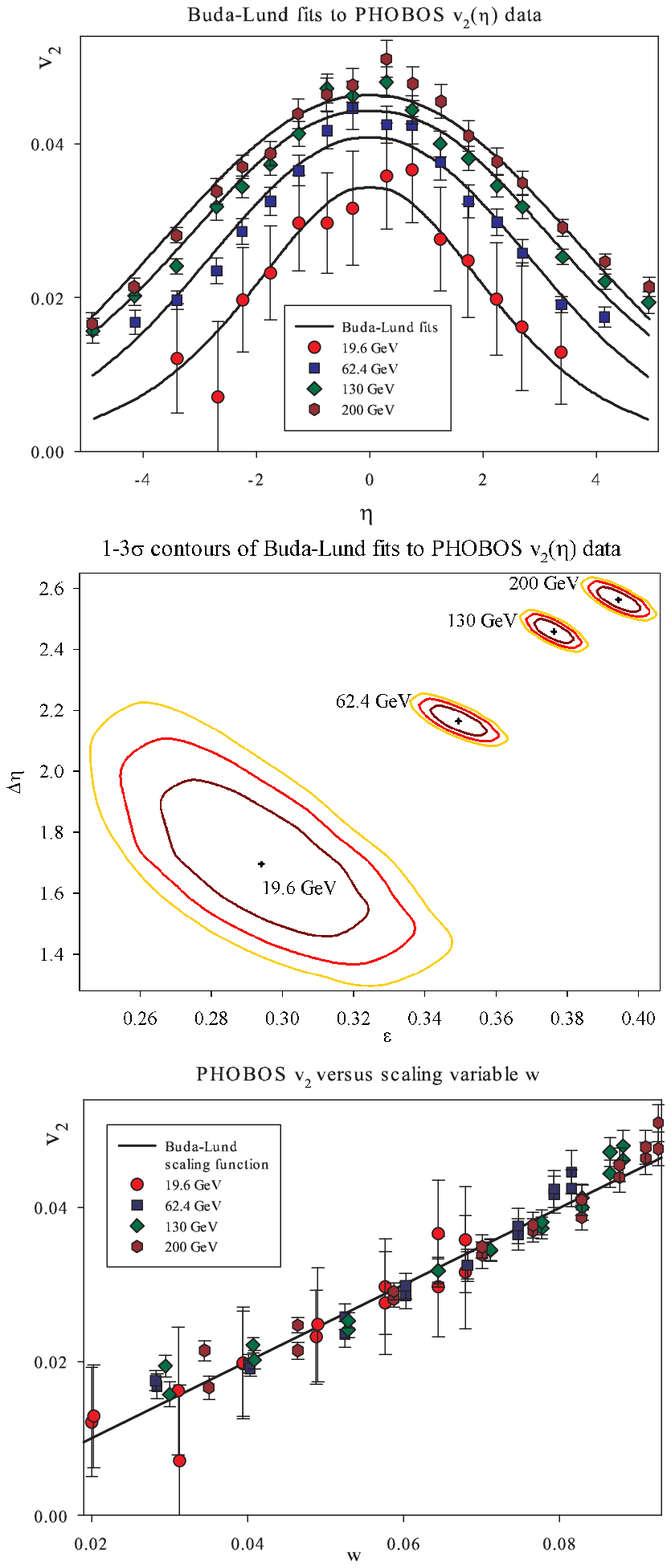}\\
  \caption{Top: PHOBOS data on the pseudorapidity dependence of the elliptic flow~\cite{Back:2004zg},
     at various center of mass energies, with Buda-Lund fits. Middle: Error contours of the fits.
     Bottom: Elliptic flow versus scaling variable $w$ is plotted.
     The data points of ref.~\cite{Back:2004zg} show
     theoretically predicted~\cite{bl-ell}) universal scaling,
    when plotted against the universal scaling variable $w$.}
  \label{f:v2eta}
\end{figure}

Fits to PHOBOS data of ref.~\cite{Back:2004zg} and its 1-3$\sigma$ error contours
are shown on the top two panels of fig.~\ref{f:v2eta}. The fitting package is available at ref.~\cite{blcvs}.
Bottom panel of fig.~\ref{f:v2eta} demonstrates that the investigated PHOBOS data points follow the
theoretically predicted scaling law.

\begin{table}[ht]
\begin{tabular}{|l|c|c|c|c|}
\hline
                  & 19.6 GeV & 62.4 GeV & 130 GeV & 200 GeV \\
\hline
  $\epsilon$ & 0.294 $\pm$ 0.029  & 0.349 $\pm$ 0.008  & 0.376 $\pm$ 0.005  & 0.394 $\pm$ 0.006 \\
  $\Delta \eta$ & 1.70  $\pm$ 0.25   & 2.16  $\pm$ 0.05   & 2.46  $\pm$ 0.04   & 2.56  $\pm$ 0.04  \\
  $\chi^2$/NDF  & 1.84/11          & 20.1/13         & 34.8/15         & 27.5/15        \\
  conf. level   & 100\%           & 21.4\%          & 1.00\%          & 7.03\%         \\
\hline
\end{tabular}
  \caption{Results of fits to PHOBOS data of ref.~\cite{Back:2004zg}. Both space-time eccentricity ($\epsilon$) and longitudinal elongation ($\Delta \eta$) increase with increasing $\sqrt{s_{\textrm{NN}}}$. Remaining parameters were fixed as follows: $T_0=175$ MeV, $a=1.19$ and $u_t=1.64$.}\label{t:fit}
\end{table}

\pagebreak

\section{Rapidity dependent HBT radii}

\begin{figure}
  \includegraphics[width=500pt]{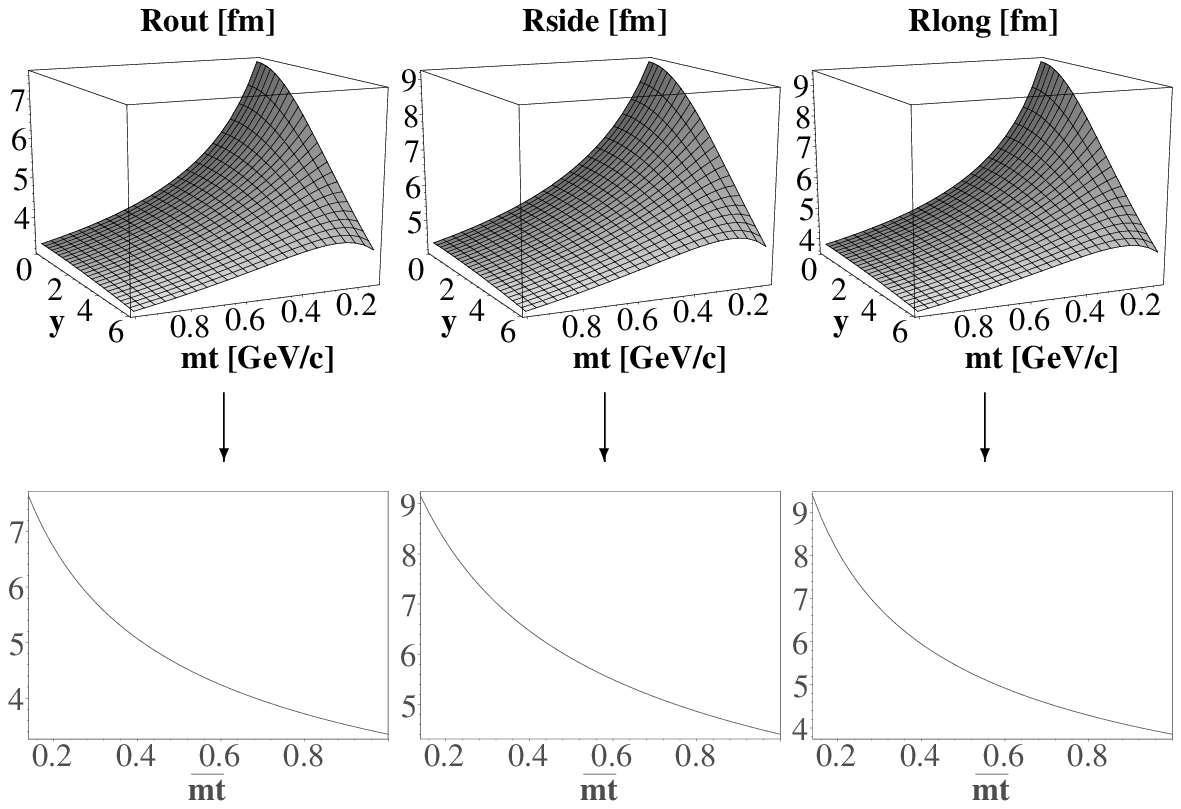}\\
  \caption{Upper panel: $R_\textrm{out}$, $R_\textrm{side}$ and $R_\textrm{long}$ as a function of
           rapidity $y$ and transverse mass $m_t$. Lower panel: The two-dimensional $R(m_t,y)$ functions
	   are predicted to show a scaling behavior, insofar as they depend only on scaling variable $\overline{m_t}$.}
  \label{f:HBTeta}
\end{figure}

In the framework of the model we can also calculate the HBT radii.
In the simplest case, where system of ellipsoidal expansion equals the
out-side-longitudinal coordinate system:
\begin{eqnarray}
R_\textrm{out}^2&=&X^2\left(1+{\frac{\overline{m}_t\, \left(a^2+\dot X^2
 \right)}{T_0}}\right)^{-1}\textnormal{, } \label{e:HBT1}\\
R_\textrm{side}^2&=&Y^2\left(1+{\frac{\overline{m}_t\, \left(a^2+\dot Y^2
 \right)}{T_0}}\right)^{-1}\textnormal{, } \label{e:HBT2}\\
R_\textrm{long}^2&=&Z^2\left(1+{\frac{\overline{m}_t\, \left(a^2+\dot Z^2
 \right)}{T_0}}\right)^{-1}\textnormal{.} \label{e:HBT3}
\end{eqnarray}
This means, that the HBT radii depend on transverse mass and rapidity only through the
scaling paremeter $\overline{m_t}$, as illustrated on fig.~\ref{f:HBTeta}. This behavior
could easily be checked by measurement of rapidity and transverse momentum
dependence of the HBT radii and comparing this data to the present prediction of the Buda-Lund model.
Such a comparision could be a further test of perfect fluid hydrodynamics.

\section{Conclusions}

In summary, we have shown that the excitation function of the
pseudorapidity dependence of the elliptic flow in Au+Au collisions
is well described with the formulas that are predicted by the
Buda-Lund type of hydrodynamical calculations.

We have provided a quantitative evidence of the validity of the
perfect fluid picture of soft particle production in Au+Au
collisions at RHIC but also show here that this perfect fluid
extends far away from mid-rapidity.

We also suggest a further test of perfect fluid hydrodynamics at
large rapidities, expressed by
eqs.~\ref{e:HBT1}-\ref{e:HBT3} and illustrated by fig.~\ref{f:HBTeta}.

The universal scaling of PHOBOS $v_2(\eta)$, expressed by
eq.~\ref{e:v2w} and illustrated by fig.~\ref{f:v2eta}
provides a successful quantitative as well as qualitative test for the
appearence of a perfect fluid in Au+Au collisions at various
colliding energies at RHIC.

\begin{theacknowledgments}
It is our pleasure to acknowledge the inspiring discussions
with Roy Lacey, Stephen Manly and G\'abor Veres.
This research was supported by the NATO Collaborative Linkage Grant PST.CLG.980086,
by the Hungarian - US MTA OTKA NSF grant INT0089462
and by the OTKA grant T038406 .
\end{theacknowledgments}

\bibliographystyle{aipprocl} 

\begin{thebibliography}{19}
\bibitem{Back:2004zg}
  B.~B.~Back {\it et al.}  [PHOBOS Collaboration],
  Phys.\ Rev.\ Lett.\  {\bf 94}, 122303 (2005)

\bibitem{Adler:2003kt}
  S.~S.~Adler {\it et al.}  [PHENIX Collaboration],
  Phys.\ Rev.\ Lett.\  {\bf 91}, 182301 (2003)

\bibitem{Adler:2001nb}
  C.~Adler {\it et al.}  [STAR Collaboration],
  Phys.\ Rev.\ Lett.\  {\bf 87}, 182301 (2001)

\bibitem{Sorensen:2003wi}
  P.~Sorensen  [STAR Collaboration],
  J.\ Phys.\ G {\bf 30}, S217 (2004)

\bibitem{Adcox:2004mh}
  K.~Adcox {\it et al.}  [PHENIX Collaboration],
  Nucl.\ Phys.\ A {\bf 757}, 184 (2005)

\bibitem{Adams:2005dq}
  J.~Adams {\it et al.}  [STAR Collaboration],
  Nucl.\ Phys.\ A {\bf 757}, 102 (2005)

\bibitem{Grassi:2005pm}
  F.~Grassi, Y.~Hama, O.~Socolowski and T.~Kodama,
  J.\ Phys.\ G {\bf 31}, S1041 (2005).


\bibitem{hidro3}
  T.~Cs\"org\H{o}, L.~P.~Csernai, Y.~Hama and T.~Kodama,
  Heavy Ion Phys.\ A {\bf 21}, 73 (2004)

\bibitem{Csorgo:1995bi}
T.~Cs\"org\H{o} and B.~L\"orstad,
Phys.\ Rev.\ C {\bf 54} (1996) 1390

\bibitem{ster-ismd03}
M. Csan\'ad, T. Cs\"org\H{o}, B. L\"orstad, A. Ster,
Act. Phys. Pol. B{\bf 35}, 191 (2004)

\bibitem{mate-warsaw03}
M.~Csan\'ad, T.~Cs\"org\H{o}, B.~L\"orstad and A.~Ster,
Nukleonika {\bf 49}, S45 (2004)

\bibitem{hidro1}
  S.~V.~Akkelin, T.~Cs\"org\H{o}, B.~Luk\'acs, Y.~M.~Sinyukov, M.~Weiner,
  PLB {\bf 505}, 64 (2001)

\bibitem{hidro2}
  T.~Cs\"org\H{o},
  hep-ph/0111139.

\bibitem{hidro4}
  T.~Cs\"org\H{o} and J.~Zim\'anyi,
  Heavy Ion Phys.\  {\bf 17}, 281 (2003)

\bibitem{bl-ell}
M.~Csan\'ad, T.~Cs\"org\H{o} and B.~L\"orstad,
  Nucl.\ Phys.\ A {\bf 742}, 80 (2004)

\bibitem{Kharzeev:2001gp}
  D.~Kharzeev and E.~Levin,
  Phys.\ Lett.\ B {\bf 523}, 79 (2001)
  [arXiv:nucl-th/0108006].

\bibitem{blcvs}
\verb|http://www.phenix.bnl.gov/viewcvs/offline/analysis/budalund/|


\end{thebibliography}


\end{document}